# Transient Ferromagnetism in Ultrafast Phase Transitions in Perovskites under XUV Irradiation: A Comparative Study of $SrTiO_3$ and $KTaO_3$

Aldo Artímez Peña[1,2], Nikita Medvedev[1, 3]

1) Institute of Physics, Czech Academy of Sciences, Na Slovance 1999/2, 182 00 Prague 8, Czech Republic
2) Faculty of Nuclear Sciences and Physical Engineering, Czech Technical University in Prague, Břehová 7 115 19 Praha 1, Czech Republic
3) Institute of Plasma Physics, Czech Academy of Sciences, Za Slovankou 3, 182 00 Prague 8, Czech Republic

## Abstract

We study ultrafast structural and electronic responses of strontium titanate ($SrTiO_3$, STO) and potassium tantalate ($KTaO_3$, KTO) to intense femtosecond irradiation using the XTANT-3 multiscale code. It is found that at threshold doses of 0.7 eV/atom in STO and 0.9 eV/atom in KTO, a superionic state thermally forms with selective melting of the oxygen subsystem while metallic sublattices remain ordered. This state persists up to ~1.6 eV/atom (STO) and ~1.5 eV/atom (KTO), above which complete disorder occurs. Analysis of the transient electronic density of states suggests that the B-site *d*-orbitals govern the divergent behaviour of the two materials: the compact Ti 3*d* orbitals in STO produce a narrow conduction band and large intra-atomic exchange parameter, driving a transient ferromagnetic instability on ~1 ps timescales, whereas the more spatially extended Ta 5*d* orbitals in KTO yield a broader conduction band and smaller exchange parameter, keeping KTO paramagnetic. These results suggest *d*-orbital spatial extent as a structural parameter that influences phase transition sequences, and magnetic response under extreme electronic excitation, with implications for the ultrafast optical control of electronic and magnetic properties in perovskite-based optoelectronic devices. Landau–Devonshire analysis shows that irradiation at 0.3 eV/atom transiently deepens the polar potential well in unstrained and strained STO and KTO, with the effect amplified approximately 2-fold in strained STO and 6-fold in strained KTO with respect to the unstrained cases.

# I. Introduction

Strontium titanate ($SrTiO_3$, STO) and potassium tantalate ($KTaO_3$, KTO) are prototypical perovskite oxides that have been investigated extensively due to their well-defined crystal structures and rich optoelectronic behavior[1–4]. Both adopt the cubic $Pm\overline{3}m$ perovskite structure at room temperature and share broadly similar equilibrium electronic architectures — with the top of the valence band dominated by O 2*p* states and the bottom of the conduction band derived from the $t_{2g}$ manifold of Ti 3*d* and Ta 5*d* states, respectively[5–10]. Both materials have attracted considerable attention in recent years as promising candidates for next-generation photovoltaic and optoelectronic applications due to their pronounced dielectric polarization, thermal and chemical stability, and comparatively low fabrication costs[11,12].

The interaction of laser radiation with these materials has been widely explored as a means of tailoring their structural, electronic, and magnetic properties[13–16]. However, the majority of studies focused on irradiation conditions characterized by relatively low dose rates, leaving the response of these systems to extreme, ultrafast excitation less thoroughly understood. More recently, X-ray free-electron laser (XFEL) experiments on STO-containing heterostructures have shown that hard X-ray pulses can induce structural damage in STO[17].

In general, laser–matter interaction proceeds through a sequence of processes[18,19]. Initially, incident photons are absorbed by electrons, promoting them to higher-energy states. In semiconductors and insulators, this typically involves transitions from the valence band to the conduction band. At higher photon energies, such as those delivered by extreme ultraviolet (XUV) or x-ray sources, core-level photoionization becomes the dominant absorption channel. This process generates highly energetic photoelectrons and leaves core holes, which predominantly relax *via* Auger decay on femtosecond timescales, while radiative recombination remains comparatively inefficient for light elements and outer shells[20,21].

The primary photoexcited electrons subsequently initiate secondary electron cascades through processes such as impact ionization, plasmon excitation, and electron–phonon (electron-ion) scattering.[18,21] These interactions rapidly redistribute energy within the electronic system, leading to the establishment of a quasi-equilibrium (Fermi–Dirac) distribution typically within sub-picosecond timescales.

Energy transfer from the electronic system to the lattice occurs through two principal mechanisms. The first involves nonadiabatic electron–phonon coupling, which governs lattice heating on picosecond timescales (thermal effects). The second mechanism arises from transient modifications of the interatomic potential energy surface due to electronic excitation, which occur on sub-picosecond timescales (nonthermal effects).

Under sufficiently high excitation densities, these effects can induce nonthermal structural transformations, including ultrafast melting, even before significant lattice heating[22,23]. Such highly nonequilibrium conditions can give rise to transient states inaccessible within the equilibrium phase diagram[24,25].

Free-electron lasers provide a unique platform for accessing this regime, delivering intense femtosecond pulses in the XUV and x-ray spectral ranges[26–29]. The combination of ultrashort pulse duration, high photon flux, and homogeneous volumetric heating results in extremely high dose rates, enabling excitation pathways that differ fundamentally from those observed under steady-state or low-dose irradiation[29,30]. These conditions facilitate the generation and investigation of strongly nonequilibrium states of matter and offer new opportunities for controlled modification of material properties on ultrafast timescales[31].

In this work, we investigate the response of STO and KTO to intense femtosecond XUV irradiation using the XTANT-3 computational framework[32]. With focus on identifying damage thresholds and the mechanisms governing structural modification, characterizing the transient states — including superionic states, melting, and ferromagnetism — that emerge under these extreme excitation conditions, and assessing the potential for transiently induced magnetic ordering driven by irradiation-modified electronic structure.

## II. Model

The effects of ultrafast irradiation on KTO and STO in the $Pm\overline{3}m$-phase are investigated using the XTANT-3 combined code[32]. The code integrates several interdependent simulation modules that exchange information on-the-fly, enabling a self-consistent treatment of coupled electronic and atomic dynamics[33]. Its components directly relevant to the present study are outlined below; further details can be found in the corresponding XTANT-3 documentation[34].

Photoexcitation, together with subsequent evolution of excited electrons, and core-hole Auger decay processes are described using event-by-event transport Monte Carlo (MC) simulation[33,35,36]. The photoabsorption cross sections, Auger lifetimes, and orbital ionization potentials are taken from the EPICS2025 database[37].

Electron cascades are followed until their energy drops below a cutoff of 10 eV relative to the bottom of the conduction band. Inelastic scattering, including impact ionization and interactions with valence electrons, is treated within the linear response theory using the Ritchie–Howie oscillator model for the complex dielectric function [38], applied in the single-pole approximation[39]. Elastic scattering of high-energy electrons is described by the Rutherford cross section with a modified Molière screening parameter[35]. Statistical convergence is ensured by averaging over 50,000 MC iterations[33,40].

The electronic distribution function defines transient populations of states in the valence and conduction bands for electrons below the cutoff energy. Scattering processes of those electrons are described using Boltzmann collision integrals that account for both electron–electron and electron–phonon (electron-ion) interactions[41]. Electron-electron scattering is described with the relaxation-time approximation. In this work, the characteristic time is set to zero (instantaneous relaxation toward Fermi-function). Previous studies indicate that under strong excitation conditions, deviations from instantaneous thermalization have only a limited impact on the predicted structural response, typically modifying damage thresholds by no more than 15%[41].

Electron–phonon scattering (nonadiabatic interaction) is treated using the dynamical coupling approach, where the coupling matrix elements are obtained from the evolving tight-binding (TB) Hamiltonian[42]. The electronic structure and interatomic forces are calculated within a transferable tight-binding framework[43]. The nonorthogonal TB Hamiltonian depends explicitly on the instantaneous coordinates of all atoms in the simulation box. Its diagonalization at each timestep provides the transient electronic states and the corresponding potential energy surface. Changes in the electronic distribution induced by irradiation, therefore, directly modify the interatomic forces, enabling the description of nonthermal phase transitions[36,44]. For both materials, the periodic table baseline parametrization (PTBP), based on an $sp^3d^5$ LCAO basis set, is used[45,46].

Atomic response is simulated using molecular dynamics (MD), with forces derived from the tight-binding potential energy surface. Energy transferred from electrons is delivered to atoms *via* a velocity scaling procedure applied at each timestep[42]. Atomic trajectories are propagated using the fourth-order Martyna–Tuckerman algorithm with a timestep of 0.1 fs in the microcanonical (NVE) ensemble[47].

Simulations are performed for 320-atom supercells for both KTO and STO, constructed from the unit cells available in Ref. [48] and subsequently relaxed using the steepest descent method, yielding equilibrium densities of 6.01 g/cm$^3$ and 4.22 g/cm$^3$, respectively (experimental studies report slightly higher values of 6.95-7.02 g/cm$^3$ for KTO[49] and 5.11-5.12

g/cm$^3$ for STO[50], which is typical for transferrable tight binding methods). These system sizes are sufficient for reliable modelling [36]. Periodic boundary conditions are applied to mimic bulk material.

Atomic velocities are initialized according to the Maxwell–Boltzmann distribution at room temperature, followed by a 200-fs equilibration period prior to irradiation. The simulations are then continued up to 25 ps after the pulse centred at 0 fs[51]. Structural evolution is analysed using the OVITO visualization package[52].

The total and partial electronic density of states (DOS), electronic heat capacity, and electronic heat conductivity are evaluated using a 7×7×7 Monkhorst–Pack k-point grid over the supercell[53]. The electron heat capacity is determined from the temperature derivative of the electronic entropy, while the corresponding chemical potential is obtained numerically from the transient electronic energy density[54]. Electronic thermal conductivity is evaluated from the Onsager transport coefficients, accounting for both electron–phonon and electron–electron scattering contributions[55].

In contrast, the electron–phonon coupling strength is evaluated using a nonperturbative dynamical coupling approach that derives the energy transfer rates directly from time-dependent simulations varying electronic temperature [42].

The XTANT-3 framework has previously been validated through comparisons with experimental damage thresholds across a range of materials, showing reasonable agreement[25,36,56,57]. In addition, consistency with density functional theory calculations within the Born–Oppenheimer approximation was demonstrated[25], while time-dependent DFT studies supported the nonadiabatic effects captured by the model[58].

## III. Results

We start by analyzing the thermodynamic properties of the studied perovskites, proceeding with the analysis of their dynamical response to radiation. The damage thresholds are then identified, and the evolution of the magnetic state is estimated.

### A. Thermodynamic properties

The calculated electronic density of states (DOS) in perfect crystal KTO and STO are shown in **Figure 1**. The DOS of the pristine systems is in reasonable agreement with previously reported calculations[59–62]. The TB-calculated peaks in both materials are narrower than the DFT-calculated ones, which is a consequence of the lower equilibrium densities. However, the calculated band gaps in the pristine state are ~5.5 eV for KTO and ~4.5 eV for STO, only slightly overestimating the experimental values (4.35-5.03 eV for KTO[63] and 3.70-3.78 eV for STO[2]), validating the simulations set up. Keeping in mind these discrepancies, further calculated results may be interpreted carefully.

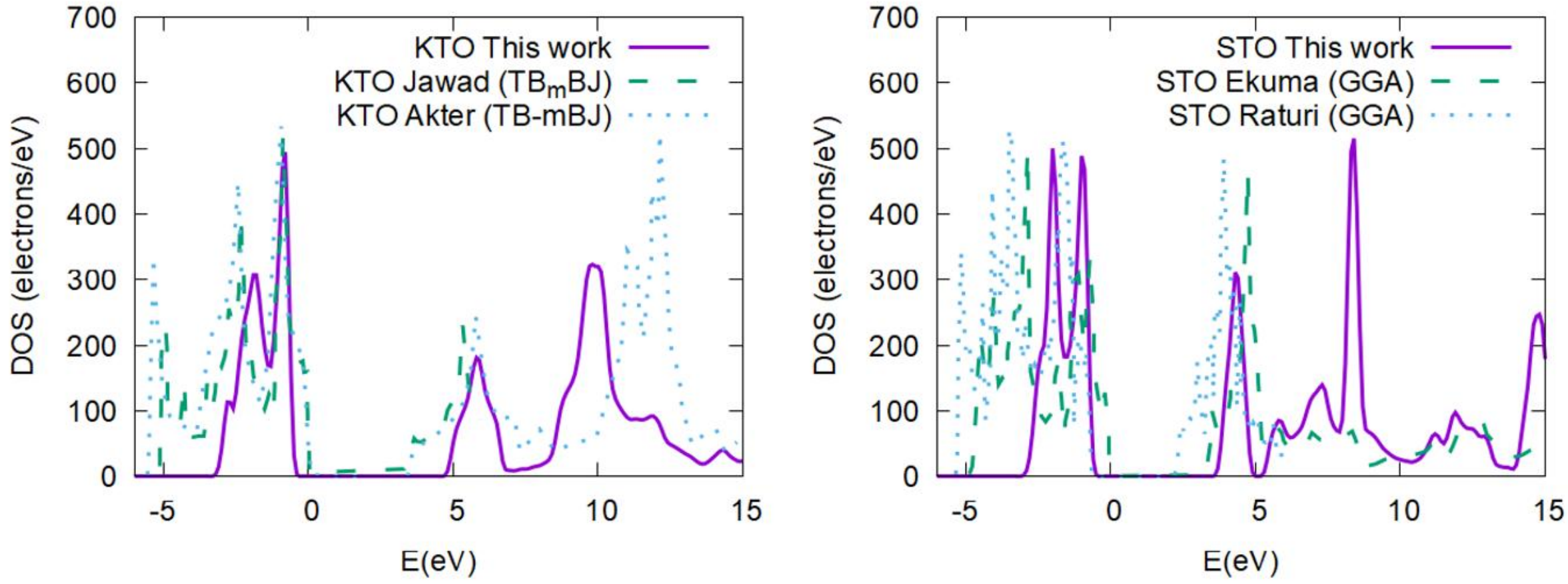

***Figure 1.*** *Total electronic density of states in pristine KTO (left panel) and STO (right panel) calculated with XTANT-3. DFT within GGA[61,62] and TB[59,60] calculations are shown for comparison.*

For evaluation of the material response to laser irradiation, two-temperature model (TTM) or its derivatives are often used. For the description of the electronic system, the electron-temperature-dependent electronic heat capacity, heat conductivity, and electron-phonon coupling parameters are required[18,19].

The electronic heat capacity and heat conductivity, calculated with XTANT-3 in the two perovskites, are shown in **Figure 2**. The shape is typical for semiconductors [64], exhibiting very small values at the electronic temperatures significantly smaller than the band gap, rising sharply at higher temperatures. The electronic heat conductivity reaches a plateau at the temperatures of ~1 eV, suppressed by the electron-electron contribution (as seen from the fact that the electron-phonon one keeps increasing, see **Figure 2**).

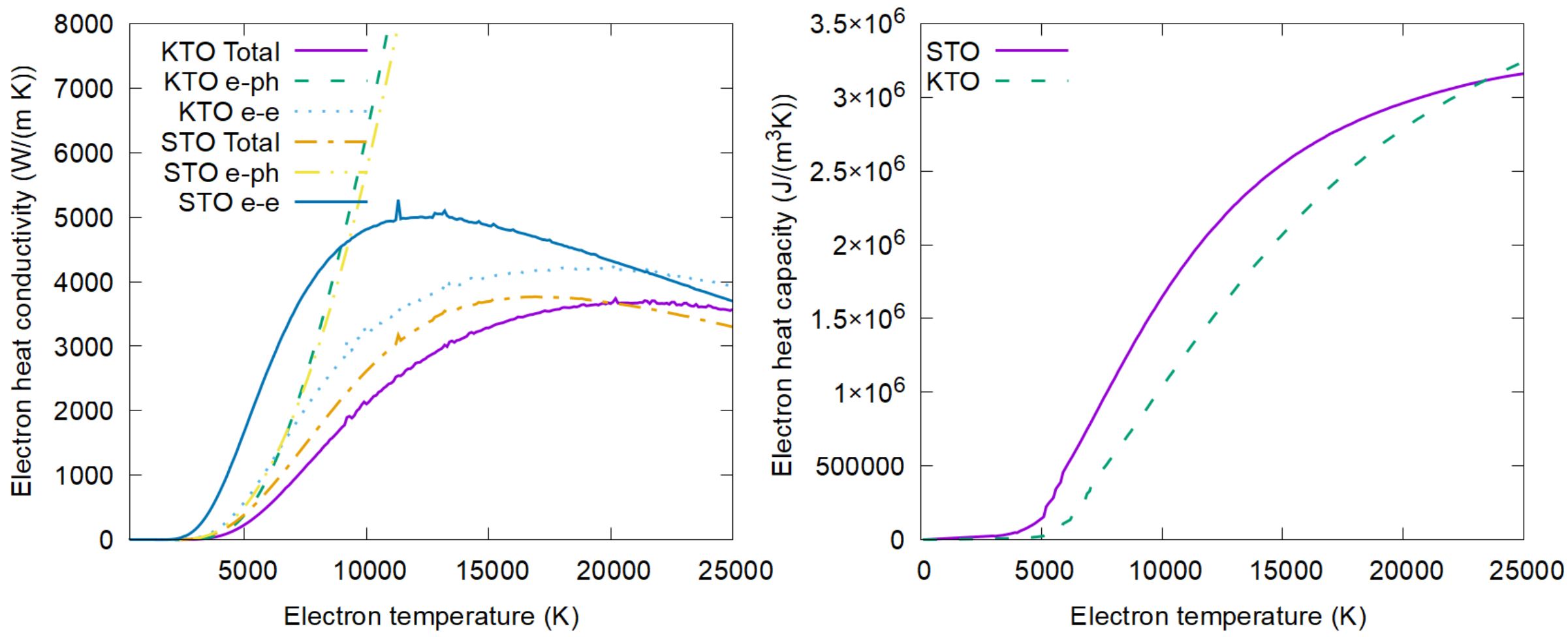

***Figure 2.*** *Total, electron-phonon and electron-electron contributions of electron heat conductivity (left panel) and electron heat capacity (right panel) in KTO and STO calculated with XTANT-3.*

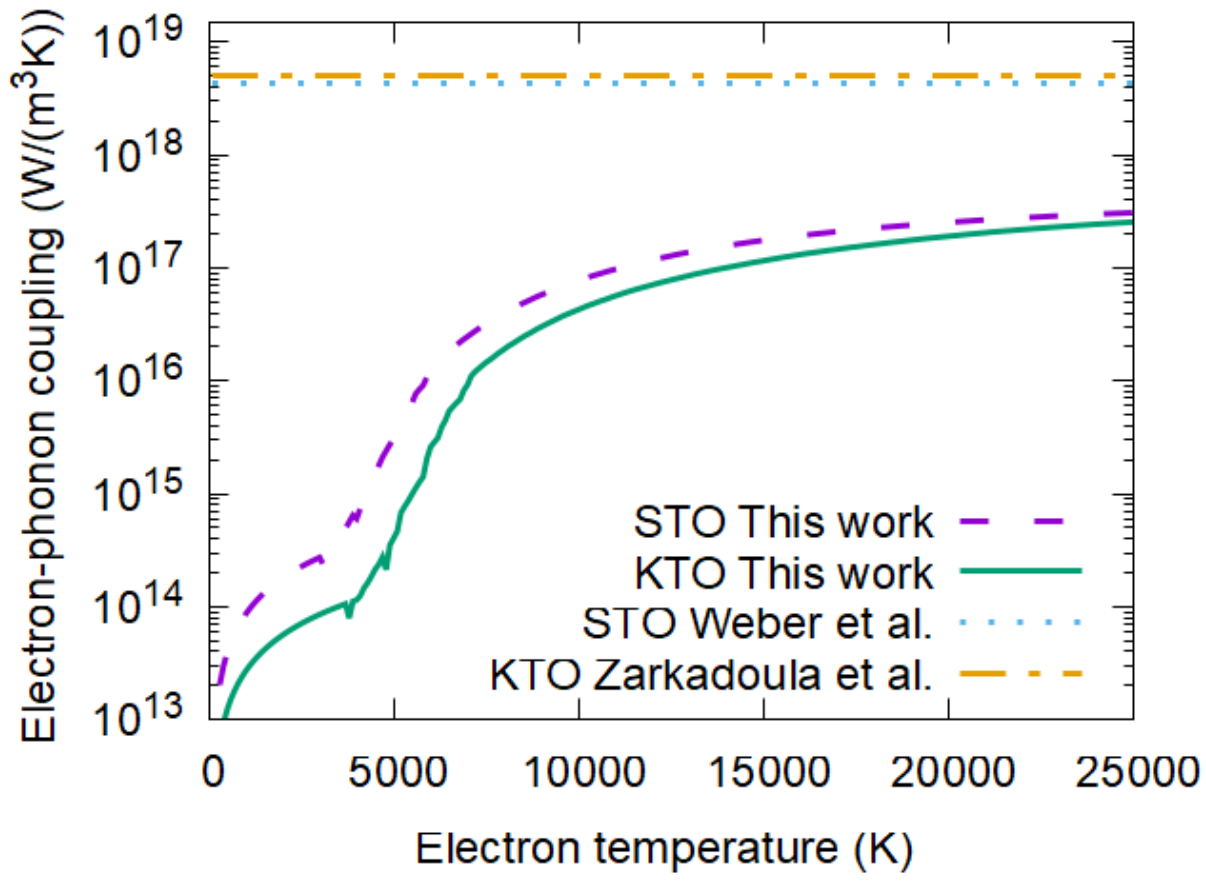

***Figure 3.*** *Electron-temperature-dependent electron-ion coupling parameter in KTO and STO; average values reported in Ref.* [65] *and* [66] *are shown for comparison.*

**Figure 3** shows the dependence of the electron-phonon coupling parameter on the electronic temperature; the values are in the same order of magnitude as in other semiconducting oxides[64]. The coupling parameter in STO is higher than in KTO, which aligns with the notion of decreasing coupling with the increase in the atomic mass, previously reported in other materials[64,67].

Values reported by Weber et al.[66] and Zarkadoula et al.[65] are an order of magnitude higher than the maximum values presented in this work. Refs. [[66],[65]] assumed a constant coupling parameter, extracting it by fitting the two-temperature model calculated melting radius to the experimental value of a swift-heavy ion track [65,66,68]. However, it has been demonstrated that such a procedure based on ion track radii drastically overestimates the value of the coupling parameter, since it inherently attributes the nonthermal effects to the coupling; discerning them produces significantly lower coupling, as can also be seen here by comparison with our results[69].

## B. Mechanisms of damage

A sequence of simulations varying the irradiation dose from 0.1 to 6.0 eV/atom, in steps of 0.1 eV/atom, allowed us to calculate the threshold doses of multiple types of damage in both materials, summarized in **Table 1**. The first damage onsets at the doses of 0.9 eV/atom and 0.7 eV/atom in KTO and STO, respectively, as seen in **Figure 4** and **Figure 5** (and corresponding atomic snapshots in **Figure 15** and **Figure 16** in the Appendix). At these threshold doses, the oxygen sublattice in both materials starts diffusing while the displacement of the metallic sublattices saturates. This is a hallmark of superionic states in which materials simultaneously exhibit solid and liquid sublattices[24,25,70].

This behaviour is observed within a range of doses up to 1.5 eV/atom in KTO and 1.6 eV/atom in STO. At higher doses, both systems disorder completely.

***Table 1.*** *Threshold doses for different damage mechanisms in KTO and STO calculated with XTANT-3*

| Damage mechanism | KTO | STO |
|---|---|---|
| Superionic state (nonadiabatic) | 0.9 eV/atom | 0.7 eV/atom |
| Melting (nonadiabatic) | 1.5 eV/atom | 1.6 eV/atom |
| Melting (adiabatic) | 3.5 eV/atom | 6.0 eV/atom |
| Band gap collapse (nonadiabatic) | 2.0 eV/atom | 3.0 eV/atom |
| Band gap collapse (adiabatic) | 3.6 eV/atom | 6.0 eV/atom |

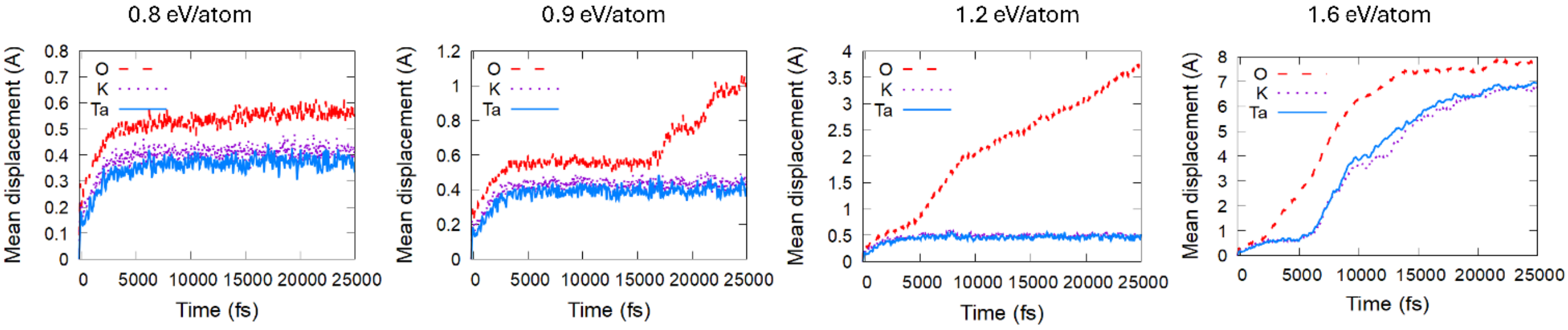

***Figure 4.*** *Element-resolved mean displacement in KTO irradiated with a 10-fs Gaussian pulse of 30 eV photon energy at different doses.*

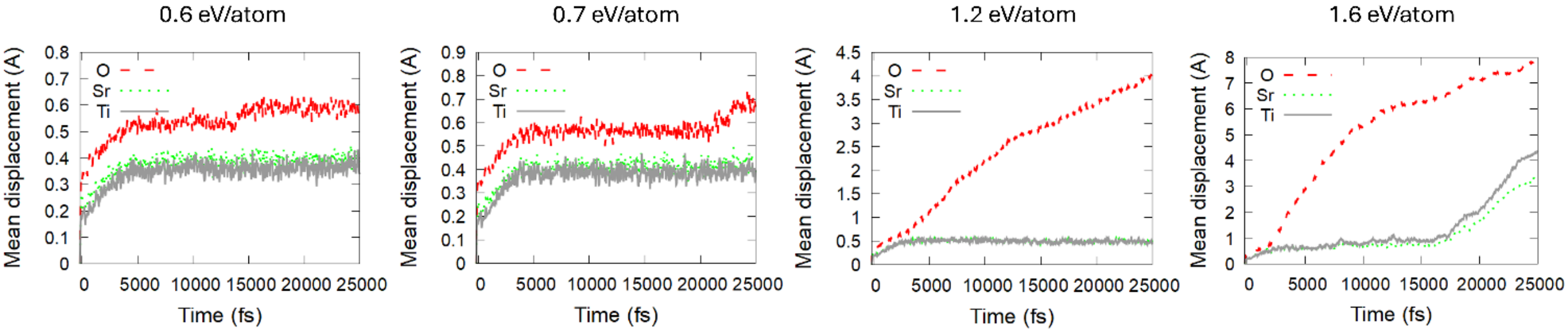

*__Figure 5.__ Element-resolved mean displacement in STO irradiated with a 10-fs Gaussian pulse of 30 eV photon energy at different doses.*

The transition toward disorder through superionic states can be qualitatively understood by examining the evolution of the material's density of states. **Figure 6** and **Figure 7** present the evolution of the DOS under irradiation at 1.2 eV/atom and 2.0 eV/atom in order to capture both the superionic-to-molten progression and the dose range over which the two materials' band-gap behavior begins to diverge. Each dose is shown once under the full nonadiabatic (non-BO) treatment and once under the Born–Oppenheimer approximation.

In the initial state, the top of the valence band is primarily composed of oxygen *p*-states. As the electronic temperature increases, electrons are progressively excited from the top of the valence band to the bottom of the conduction band. This weakens the bonds of O atoms, upon a certain dose, leading to their detachment[71,72]. Sufficiently sparse atomic structure allows for their diffusion within the solid matrix of other elements, creating a transient superionic state.

Born-Oppenheimer (BO) simulations (which artificially exclude the electron-phonon coupling and associated atomic heating by excited electrons) show that purely non-thermal effects may also result in damage; however, at significantly higher threshold doses (cf. **Table 1**), similar to other semiconductors and oxides[64]. Therefore, transition to disorder arises from an interplay of nonadiabatic heating and nonthermal effects, with the former being the main contribution[73]. This can be confirmed by the evolution of the DOS after irradiation (**Figure 6** and **Figure 7**): under the BO approximation, the DOS is practically unchanged with respect to the pristine DOS at the shown doses.

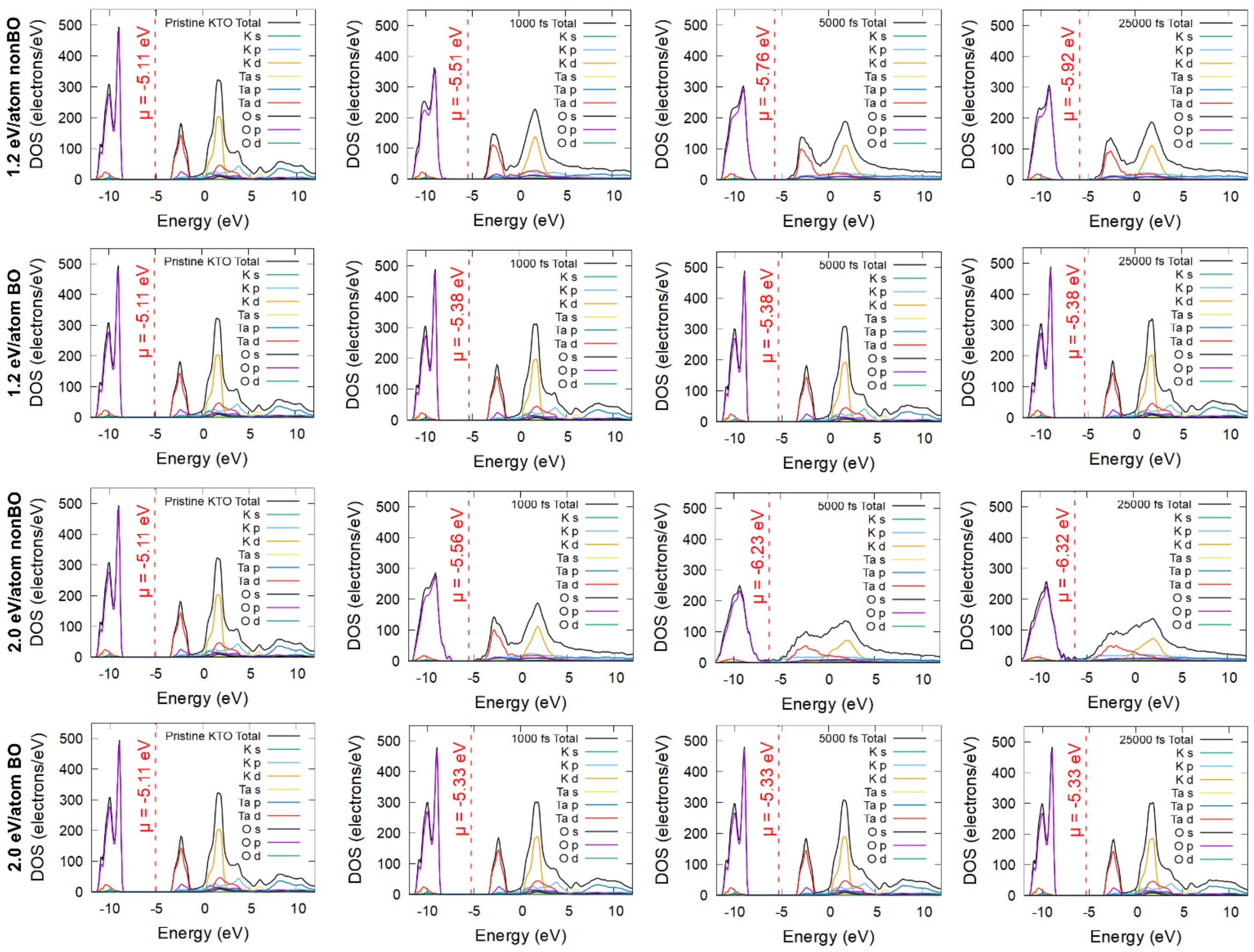

***Figure 6***. *Temporal evolution of KTO orbital-resolved PDOS and total DOS after irradiation with a 10-fs Gaussian pulse of 30 eV photon energy at various doses. Results under BO and non-BO approximations are shown. Chemical potential (μ) is marked with a vertical dashed line.*

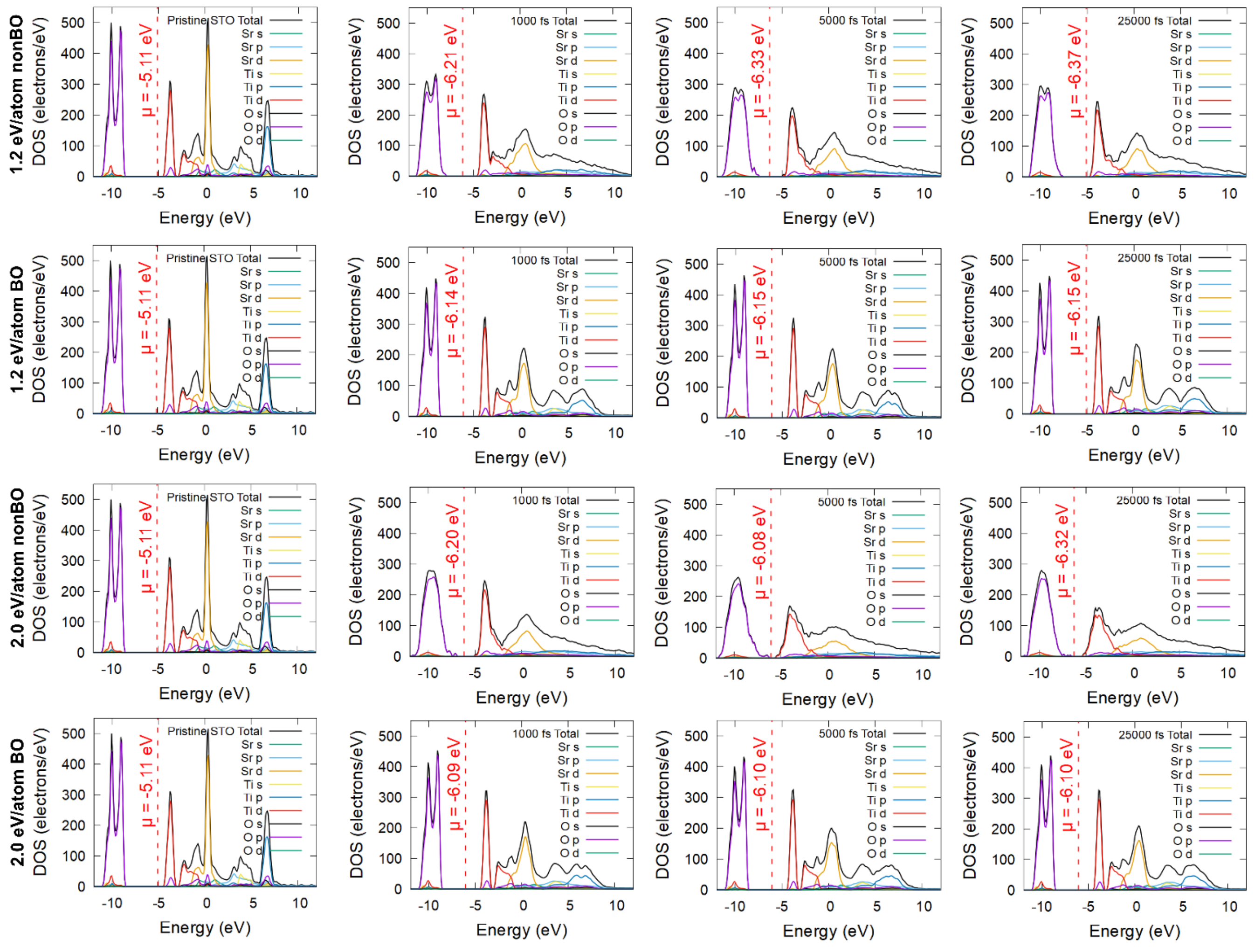

***Figure** 7. Temporal evolution of STO orbital-resolved PDOS and total DOS after irradiation with a 10-fs Gaussian pulse of 30 eV photon energy at various doses. Results under BO and non-BO approximations are shown. Chemical potential (μ) is marked with a vertical dashed line.*

We conclude that both superionic behavior and melting are mainly thermal at the threshold doses[67]; the atomic heating is driven by energy transfer from the excited electronic system *via* electron-phonon coupling (see **Figure 8** and **Figure 9**)[24,25,70].

The coupling parameter reaches its peak of ~2.5x$10^{17}$ W/($m^3$K) in STO and ~2.0x$10^{17}$ W/($m^3$K) in KTO at ~2 ps after the pulse, when the electronic temperature is still relatively high, and the atomic temperature is also close to its maximum. Afterwards, the coupling parameter decreases with a decrease in the electronic temperature. This emphasizes the importance of accounting for the dependence of the coupling parameter on both the electronic and also the atomic temperatures in reliable simulations.

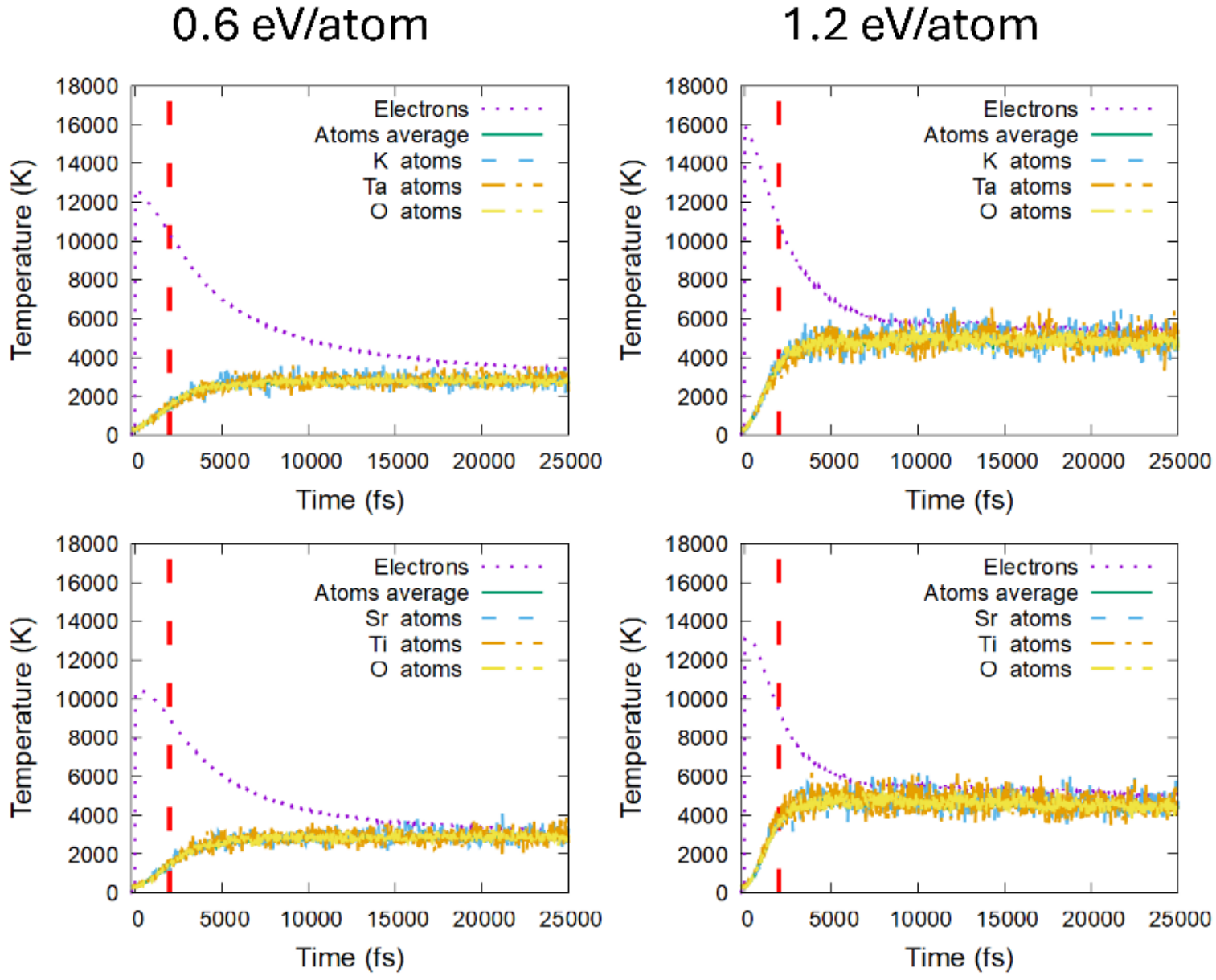

***Figure 8.*** *Electronic and atomic temperatures in KTO and STO irradiated with a 10-fs Gaussian pulse of 30 eV photon energy at different doses. The time when the electron-phonon coupling reaches its maximum is indicated by vertical dashed lines.*

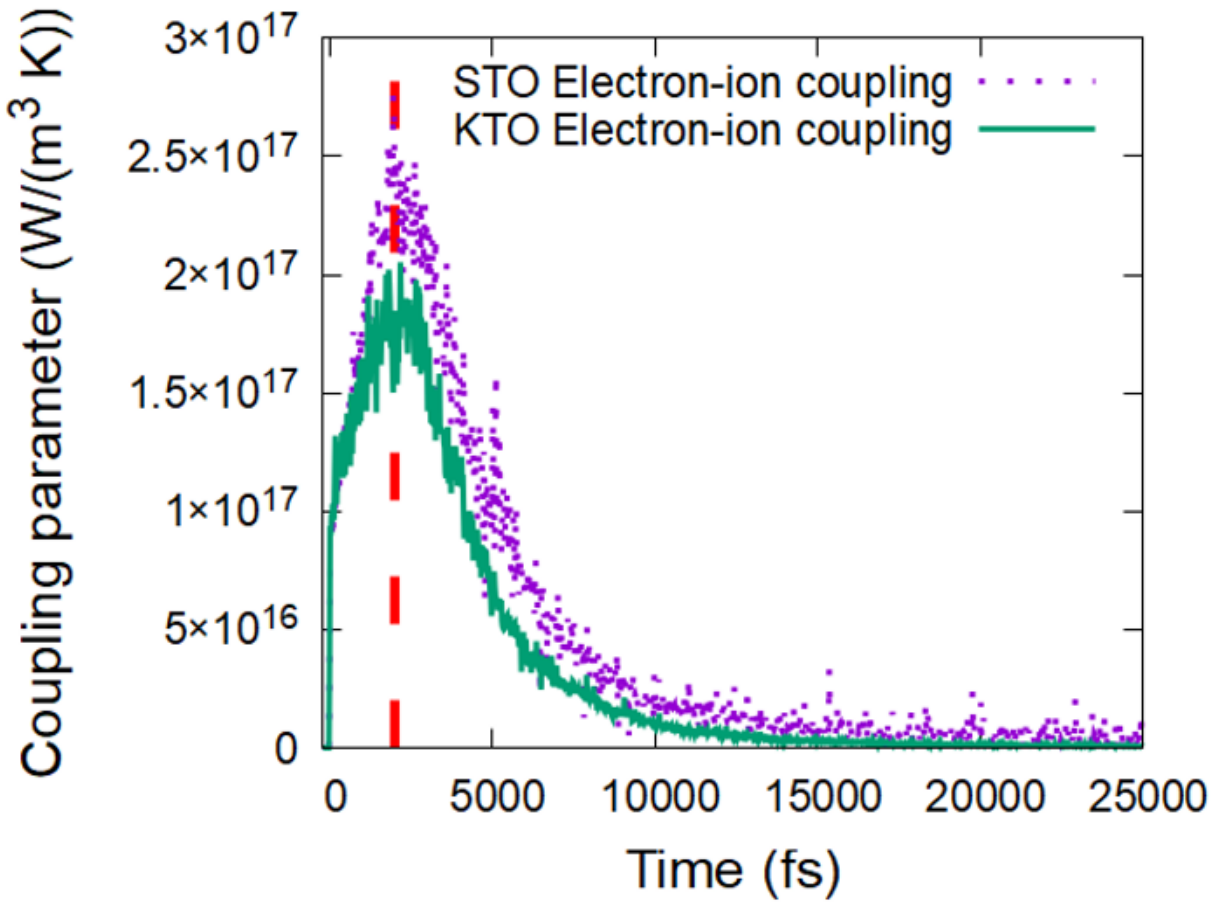

***Figure 9.*** *Time-dependent electron-ion coupling parameter in KTO and STO irradiated with a 10-fs Gaussian pulse of 30 eV photon energy at the dose of 0.6 eV/atom. The time when the electron-phonon coupling reaches its maximum is indicated by a dashed vertical line.*

In both materials under irradiation, the bandgap shrinks with the increase of the dose (**Figure 10**). KTO may transiently form semiconducting superionic states, while in STO, both superionic and melted states are semiconducting, and only above 3.0 eV/atom electronically conducting states are predicted (**Table 1**).

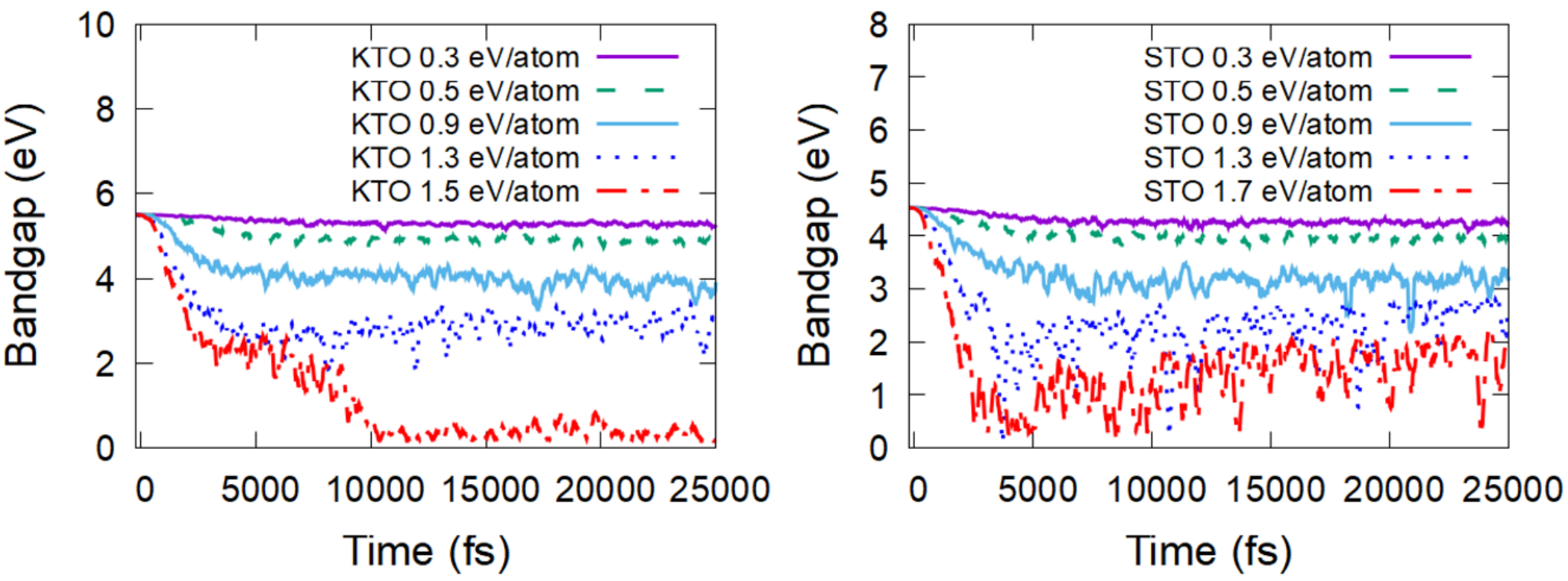

***Figure 10.*** *Temporal evolution of the bandgap of KTO and STO irradiated with a 10-fs Gaussian pulse of 30 eV photon energy at various doses.*

### C. Paramagnetic-to-ferromagnetic phase transitions

From the calculated DOS, we can obtain the Stoner number $S$ in its generalized form[74,75]:

$$S = I \cdot N_{eff,\ d}(\mu, T_e) \qquad \text{(I)}$$

Where $N_{eff,\ d}(\mu, T_e)$ is the thermally averaged d-DOS around the chemical potential (μ) at a finite electronic temperature $T_e$ and $I$ is the Stoner exchange parameter (0.45 for KTO[76,77] and 0.76 for STO[77,78]), an intra-atomic quantity that measures how strongly the exchange-correlation energy changes when spin polarization on a given atom changes[78]:

$$N_{eff}(\mu, T_e) = \int N_d(E) \cdot \left(-\frac{\partial f}{\partial E}\right) dE, \qquad \text{(II)}$$

Where $N_d(E)$ is the d-projected partial DOS at a given energy $E$, and $f = \left(e^{\frac{E-\mu}{k_B T_e}} + 1\right)^{-1}$ is the Fermi-Dirac distribution function ($k_B$ is the Boltzmann constant). The generalized definition reduces to the standard Stoner number in the limit of $T_e \to 0$ ($-\frac{\partial f}{\partial E} \to \delta(E - E_F)$ and, therefore $S = I \cdot N(E_F)$); [76,79], only applicable to metals. The Stoner number $S < 1$ indicates states in which the kinetic-energy cost of spin polarization exceeds the exchange gain arising from the spin imbalance, thereby rendering the paramagnetic state stable. In contrast, $S > 1$ is for states in which fluctuations lower the energy, and the system spontaneously polarizes[80,81].

The Stoner criterion was applied to doped wide-bandgap oxide semiconductors including CaO, BaO, and related alkaline earth oxides[82,83].

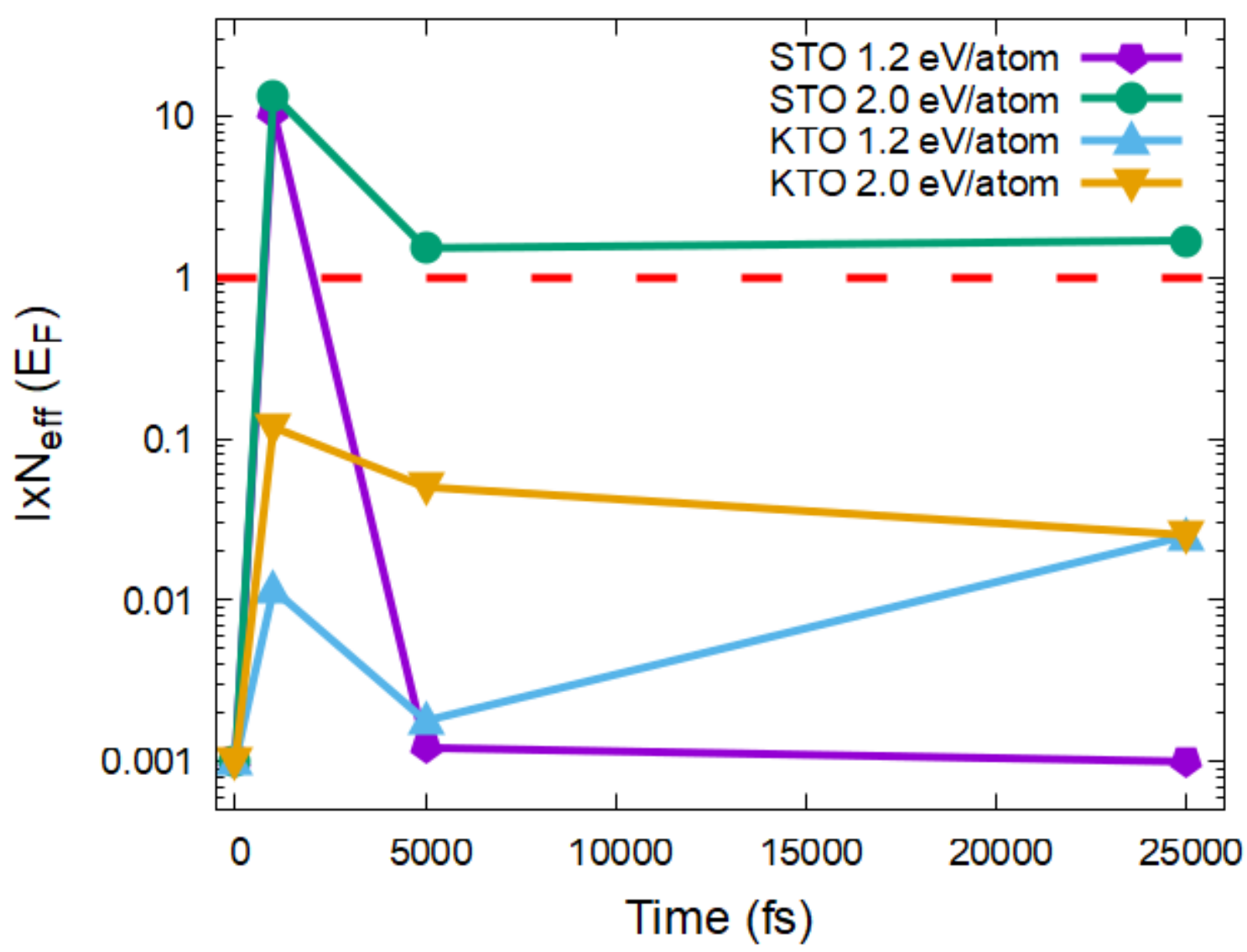

***Figure 11.*** *Temporal evolution of the Stoner number in KTO and STO after irradiation with a 10-fs Gaussian pulse of 30 eV photon energy at different doses.*

**Figure 11** shows the temporal evolution of the Stoner number in both materials under study after irradiation at various doses. For this analysis, only d-states contributions to the DOS were considered because O 2p and A-site s/p contributions to the DOS carry negligible exchange interaction[76,78].

Values rising above 1 in STO on timescales of 1 ps suggest a ferromagnetic instability. This paramagnetic-to-ferromagnetic phase transition was previously reported in STO, and it is attributed to laser and ion irradiation-induced oxygen vacancies[84–86]. Our results suggest that transient ferromagnetic states may also be achieved by ultrafast XUV irradiation. Moreover, their lifetime may be tuned with the dose.

In STO, the Ti 3*d* conduction band in the pristine material forms a narrow peak spanning above the band gap and shifts into the conduction band under irradiation-induced electron excitation (**Figure** 7.). This narrow bandwidth implies a large $N_{eff}(\mu, T_e)$ per spin channel, once electrons are promoted into this region, and the narrowness is preserved throughout the 1000–5000 fs at both 1.2 and 2.0 eV/atom doses.

In KTO, the computed *S* remains below unity throughout the simulated dose range, and the physical origin of this result is the smaller exchange parameter *I*[76,78], combined with the broader Ta 5*d* conduction band — a direct consequence of the greater spatial extent of 5*d* orbitals[87]— and this breadth is equally preserved under irradiation at all times shown. A wider band means a lower $N_{eff}(\mu, T_e)$ per spin channel. At the same time, *I* decreases as the spatial extent of the orbital increases: for compact 3*d* orbitals, the large wavefunction overlap with itself yields higher $I$ values in STO[77,78] compared to KTO[76,77]. These two factors explain why oxygen vacancies introduced by irradiation produce transient ferromagnetic instability in STO but not in KTO, supporting that the difference is of B-site $d$-band origin[84].

Let us emphasize that the above-mentioned bandgap overestimation suggests that the reported prediction of $S > 1$ in STO is a conservative estimate (a lower bound), strengthening the conclusion of ferromagnetic instability. Since $I$ is expected to be insensitive to band-structure details (given the intra-atomic character of this magnitude, determined by the self-overlap of the orbital wavefunction), the error in $N_{eff}$ associated with the tight-binding-based band structure is bound at approximately 10–20%, and 22–63% due to band-gap overestimation.

## D. Ferroelectric instability

To quantify the evolution of the ferroelectric instability under laser irradiation, we perform a Landau–Devonshire analysis of the polar order parameter from the XTANT-3[34] trajectories, computing Mulliken charges in unstrained and strained KTO and STO. Strained supercells were obtained by artificially enlarging the $c$ lattice parameter, analogous to Ref [[88]].

The Landau–Devonshire free energy expanded in even powers of the polar order parameter $Q$ takes the form $F(Q) = a_2 Q^2 + a_4 Q^4$ where $Q = \langle z_B \rangle - \langle z_A \rangle$ is the difference of the B-site and A-site mean coordinates in the direction where the strain was applied (001); using the A-site mean as a reference removes rigid-body translations. The coefficients $a_2$ and $a_4$ are determined by least-squares fitting of $F(Q)$ to the binned median of the total $\Delta E = E(t) - \langle E \rangle$ as a function of $\Delta Q = Q(t) - \langle Q \rangle$, energy and charge deviations from the mean values computed over the window $t > 5$ ps. By this time, the pulse-driven electronic excitation is expected to significantly decay. The sign of $a_2$ determines which configuration is stable[89,90]: $a_2 > 0$ gives a single minimum at $Q = 0$ – paraelectric (PE); $a_2 < 0$ produces a double-well with minima at $\pm Q_0 \neq 0$ – ferroelectric (FE); and $a_2 = 0$ is the quantum critical point (QCP)[91]. The sign of $a_4$ distinguishes a continuous second-order transition ($a_4 > 0$) from a first-order scenario ($a_4 < 0$)[90].

MD without quantum zero-point motion places quantum paraelectrics marginally on the ferroelectric side of the classical QCP[92,93]. Therefore, in the present analysis, the physically meaningful quantities are the strain-induced difference in $a_2(t)$ (**Figure 12**). KTO and STO, respectively strained at c/a = 1.150 and c/a = 1.153, show a larger pulse-induced deepening in $a_2(t)$ (by a factor of 2 for STO, and 6 for KTO) relative to the unstrained cases at the same dose, suggesting that strain shifts both systems toward the ferroelectric phase boundary, similar to the experimental observations in strained $BaTiO_3$[94]. The FE window (time during which $a_2(t) < 0$) ends at t ≈ 13–16 ps (**Figure 12**). The FE window duration across the strained systems suggests that this timescale is set by the electron–phonon thermalization (**Figure 9**).

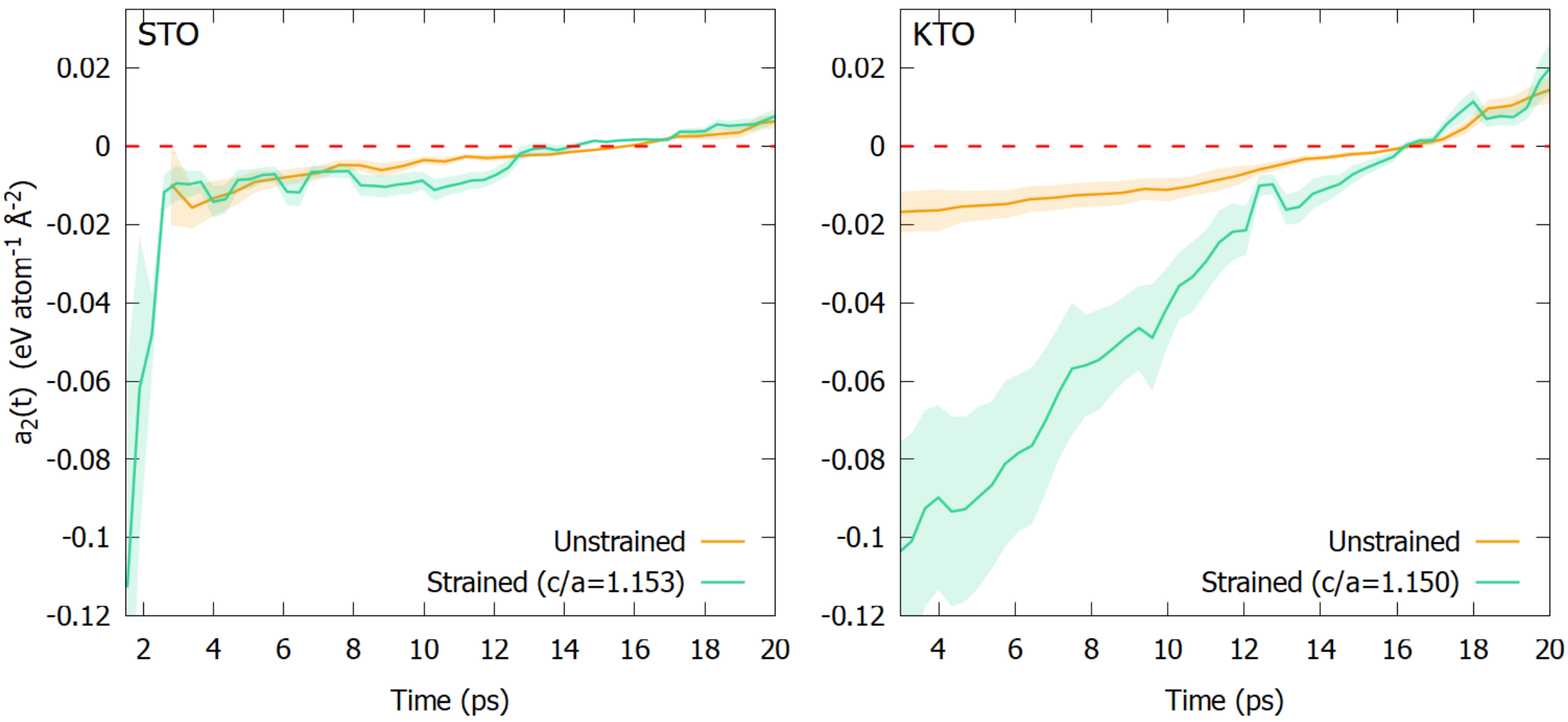

***Figure 12**. Time-resolved Landau coefficient $a_2(t)$ in unstrained and strained KTO and STO irradiated with 0.3 eV/atom. Shaded bands show $\pm\sigma$ uncertainty. The horizontal dashed line marks $a_2 = 0$ (phase boundary).*

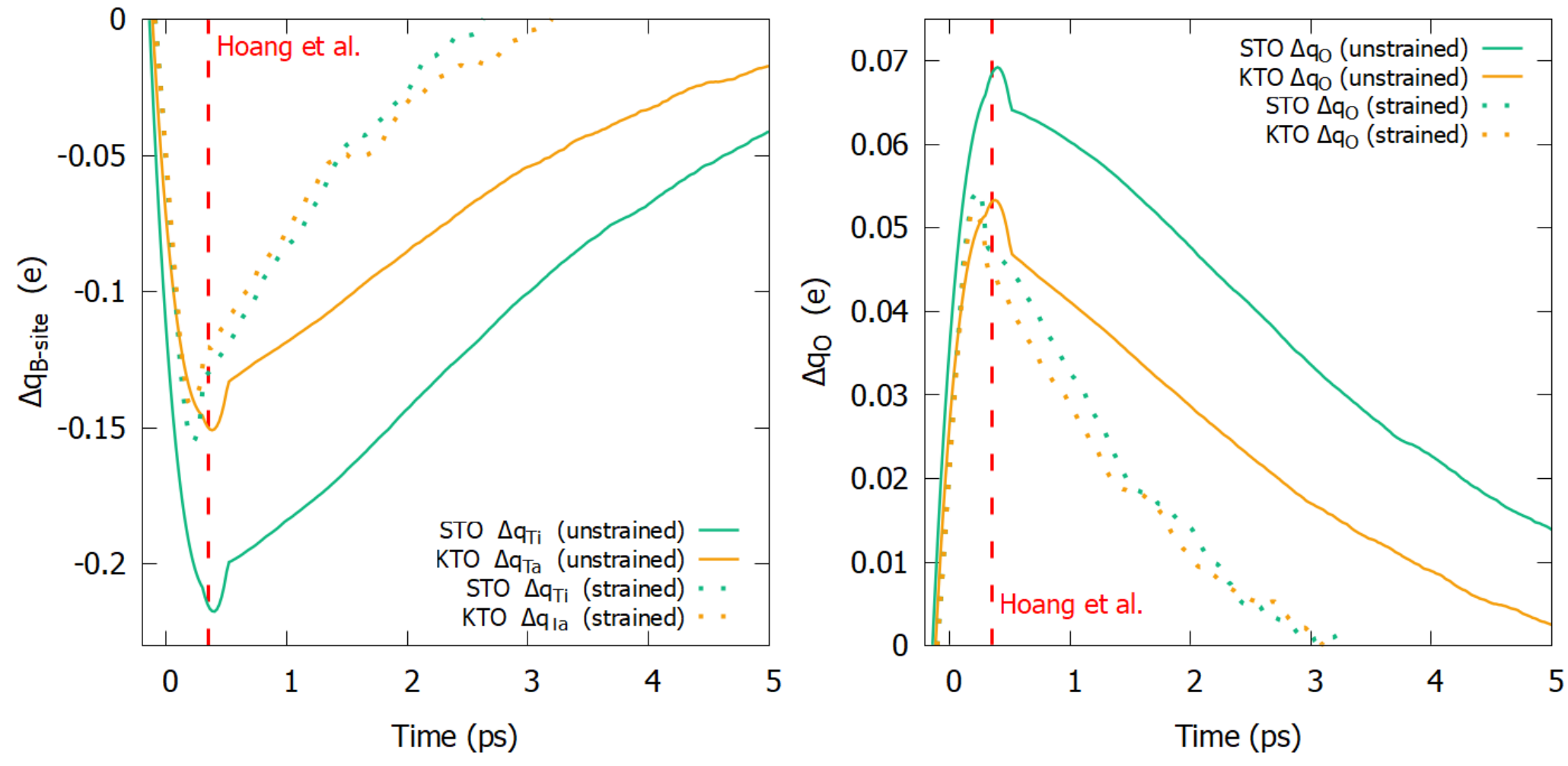

***Figure 13**. B-site cation Mulliken charge transfer $\Delta q_B(t)$ (left panel) and oxygen charge transfer $\Delta q_O(t)$ (right panel) for STO and KTO at 0.3 eV/atom, unstrained (solid) and strained (dashed). All quantities are referenced to the mean pre-pulse value. The vertical dashed line marks t=350 fs, the timescale of maximum photoexcited carrier density measured in $BaTiO_3$ by Hoang et al[95].*

**Figure 13** shows that the B-site cation undergoes a reduction in formal charge: the Ti Mulliken charge in unstrained STO drops from its pre-pulse value by $\Delta q_{Ti} = -0.218e$, peaking at t ≈ 400 fs, while in unstrained KTO the Ta charge drops by $\Delta q_{Ta} = -0.151e$, peaking at t ≈ 380 fs. Simultaneously, the oxygen Mulliken charges become less negative by $\Delta q_O = 0.069e$ (STO) and $0.053e$ (KTO). This direction and timing are consistent with the mechanism characterized by Hoang *et al.* in $BaTiO_3$ by time-resolved X-ray diffraction and second-harmonic generation[95], who measured a maximum in the photoexcited carrier density at t ≈ 350 fs. In their framework, the O-to-B-site charge transfer reduced both the local B-site charge $q_{Ti}$ and the atomic displacement $\Delta z_i$, thereby reducing the instantaneous polarization. The Mulliken charge dynamics are also consistent with the orbital repopulation mechanism proposed by Song *et al.*[88] which attributes the ultrafast modification of the ferroelectric potential to charge redistribution within the B-site *d*-manifold upon above-gap photoexcitation (see Appendix, **Figure 17**).

The theoretical framework used here is qualitatively consistent with previous works reporting the laser-induced ferroelectricity in STO[15,88,96,97] and suggests that a similar phase transition may also be achieved in KTO.

## IV. Discussion

Assuming normal photon incidence, no nonlinear effects, particle and energy transport, and no electron or photon emission from the surface, the damage threshold dose may be straightforwardly converted into the incident fluence by applying the same approach as in Ref. [36] and using EPICS2025 photoabsorption cross section for core shells and those extracted from CDF for the valence band in this conversion (see Appendix, **Figure 19**). The bulk threshold fluences in KTO and STO are shown in **Figure 14**, where reference points reported for STO by Paula *et al.*[16] and Jun Lee *et al.*[17] are shown for completeness; however, a direct comparison with these results is not possible for the following reasons.

For photon energies below the band gap, multiphoton absorption dominates the interaction of the pulse with the materials[19]; thus, the data reported by Paula *et al.* (1030 nm pulse)[16] are in the regime inaccessible to the present calculations (linear regime only, photon energies above the bandgap).

The presented damage threshold also differs substantially from that reported by Jun Lee *et al.*[17] at the photon energy of 9.7 keV. The primary reason for such a discrepancy seems to be that the target used in the experiment in Ref. [17] was a multilayered system consisting of a 25-nm-thick epitaxial $BiFeO_3$ layer on an STO substrate.

Using the specific heat for both materials $C_p$ (100.5 J/mol·K for KTO[98] and 100 J/mol·K for STO[17]), the increase in the atomic temperature $\Delta T$ within the absorption volume caused by the pulse can be estimated as $\Delta T = {}^{D}/_{C_p}$, $D$ being the deposited dose. At the respective damage threshold doses (cf. **Table 1**), the increase in temperature is $\Delta T_{KTO} = 1440\ K$, $\Delta T_{STO} = 1546\ K$, slightly below the melting point for each material (1625 K for KTO[98] and 2350 K for STO[17]), which is consistent with the reported damage mechanism: a synergistic effect of electron-phonon coupling and nonthermal effects (electronic excitation softening the interatomic potential in semiconductors).

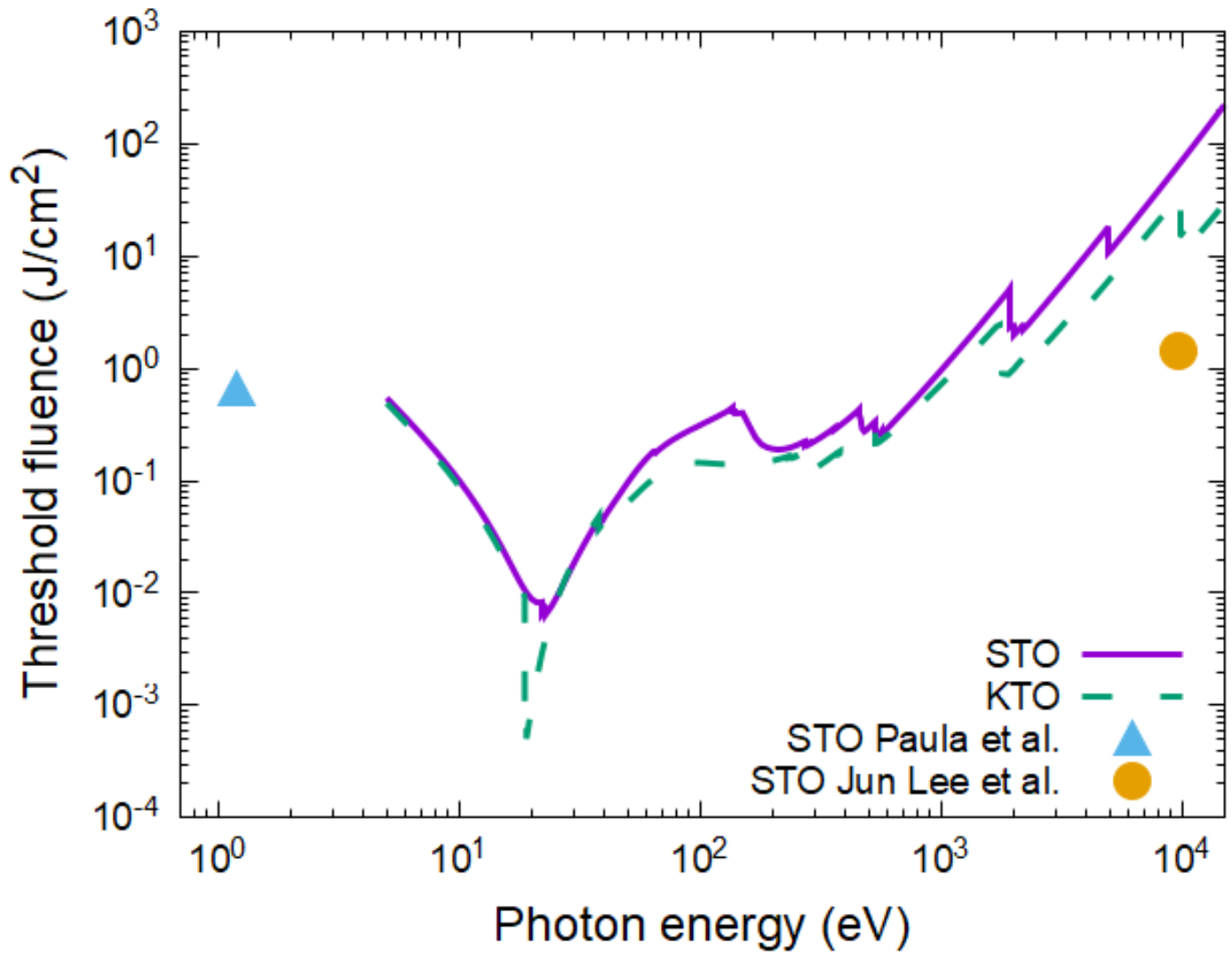

***Figure 14.*** *Damage threshold fluence in bulk KTO and STO (J/cm²) as a function of photon energies compared to the experimental data of Paula et al[16] and Jun Lee et al[17].*

Although the threshold fluences discussed above provide a useful reference for comparison with available experimental data, the simulations presented here primarily describe the ultrafast nonequilibrium response immediately following irradiation. To qualitatively examine the post-irradiation structural evolution, an additional set of simulations was performed for the deposited doses up to 5 eV/atom, allowing the systems to cool toward room temperature through the Berendsen thermostat with a characteristic relaxation time of 10 ps. This timescale was selected to ensure the full development of the transient states discussed above prior to thermal relaxation.

The inclusion of cooling leads to an increase in the effective melting damage threshold doses to 2.0 eV/atom in KTO and 2.1 eV/atom in STO. Powder diffraction patterns calculated for both materials irradiated at 1.6 eV/atom — corresponding to the melting threshold under non-equilibrium conditions (cf. **Table 1**) — show that most crystalline diffraction peaks recover by the end of the simulation (Appendix, **Figure 20** and **Figure 21**). These results suggest substantial recrystallization and high radiation resistance in both perovskite oxides, the detailed study of which is beyond the scope of the present work.

## V. Conclusions

We investigated the ultrafast structural and electronic response of two prototypical perovskite oxides, $SrTiO_3$ (STO) and $KTaO_3$ (KTO), to intense femtosecond irradiation using the simulation package XTANT-3. It is revealed that in the window of doses from 0.7 eV/atom to 1.6 eV/atom for STO and 0.9 eV/atom to 1.5 eV/atom for KTO, the oxygen sublattice begins to diffuse while the metallic sublattices remain comparatively ordered — the defining signature of a superionic state. Above those doses, both materials undergo complete melting.

Born–Oppenheimer (BO) simulations yield systematically higher threshold doses than the non-BO counterparts, confirming that at near-threshold doses, both superionic and melted states are predominantly thermal in origin, driven by energy transfer from the electronically excited system to the lattice *via* electron–phonon coupling, with nonthermal contributions playing a secondary role of softening interatomic potential (consistent with prior findings in most other semiconductors).

Analysis of the electronic density of states after irradiation suggests that with the rise in the electronic temperature, electrons are progressively promoted from O 2*p* states into the conduction band, weakening the interatomic potential of oxygen, which drives preferential oxygen diffusion and the superionic transition. With increasing excitation, further population of B-site *d* states modifies the cation potential as well, eventually producing complete melting.

These results indicate that the spatial character of the B-site *d*-orbitals determines the orbital bandwidth and inter-site hopping that controls structural stability and simultaneously governs magnetic response. In STO, the compact Ti 3*d* orbitals produce a narrow conduction band peak with large effective density of states and large exchange parameter, driving a ferromagnetic instability. In KTO, the spatially extended Ta 5*d* orbitals yield a broader band, lower effective density of states per spin channel, and smaller exchange parameter, keeping the material paramagnetic.

The *d*-orbital spatial extent thus acts as a single structural parameter that differentiates the phase transition sequences, bandgap behavior, and magnetic response of the two materials, independently of their common crystal structure and stoichiometry.

Landau–Devonshire analysis reveals that irradiation at 0.3 eV/atom transiently deepens the polar well in all systems, with an amplitude amplified 2-fold in strained STO and 6-fold in strained KTO relative to the unstrained cases. This transient modification coincides with an O 2*p* → B-site d orbital charge transfer peaking at 200–400 fs. The B-site *d*-orbital excitation identified as the electronic precursor of the polar response is also responsible for the transient paramagnetic-to-ferromagnetic instability.

Since the phase transitions identified here are governed primarily by the absorbed energy density rather than by the specific photon energy or excitation pathway, the present results are not restricted to the XUV regime but are, within the approximations of the model, transferable across a wide range of photon energies spanning from UV to hard X-rays, provided comparable energy densities are deposited on femtosecond timescales.

Both STO and KTO are of active interest as components in photovoltaic and optoelectronic architectures, and the dose-dependent tunability of the band gap, carrier density, and transient magnetic order identified here suggests that ultrafast laser excitation could serve as a non-contact handle for modifying optoelectronic response on picosecond timescales. In particular, the transient bandgap collapse in KTO and its recovery upon cooling, together with the dose-tunable ferromagnetic instability in STO, point toward potential pathways for reversible, all-optical modulation of electronic and magnetic properties in perovskite-based devices.

## VI. Conflicts of interest

There are no conflicts to declare.

## VII. Data and code availability

The code XTANT-3 used to simulate irradiation effects is available from [32].

## VIII. Acknowledgments

Computational resources were provided by the e-INFRA CZ project (ID:90254), supported by the Ministry of Education, Youth and Sports of the Czech Republic. NM thanks the financial support from the Czech Ministry of Education, Youth, and Sports (grant nr. LM2023068). The authors gratefully acknowledge the financial support from the European Commission Horizon MSCA-SE Project MAMBA [HORIZON-MSCA-SE-2022 GAN 101131245].

## IX. Appendix

**Figure 15** and **Figure 16** provide atomic snapshots of KTO and STO, respectively, at four representative doses spanning the sub-damage, superionic, and melting regimes.

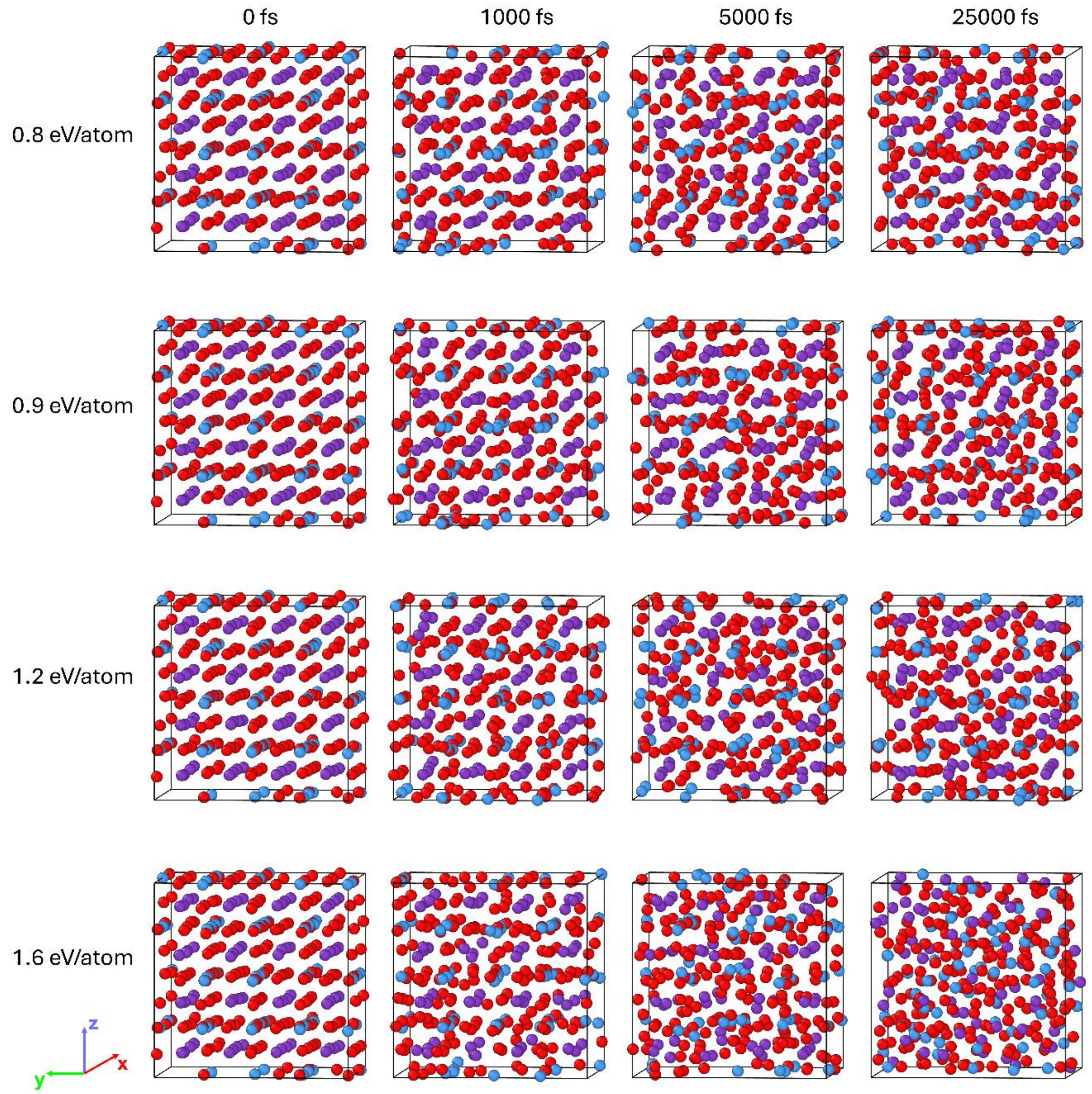

***Figure 15.*** *Atomic snapshots of KTO irradiated with a 10-fs Gaussian pulse of 30 eV photon energy at different doses. Red balls are O; violet balls are K; blue balls are Ta.*

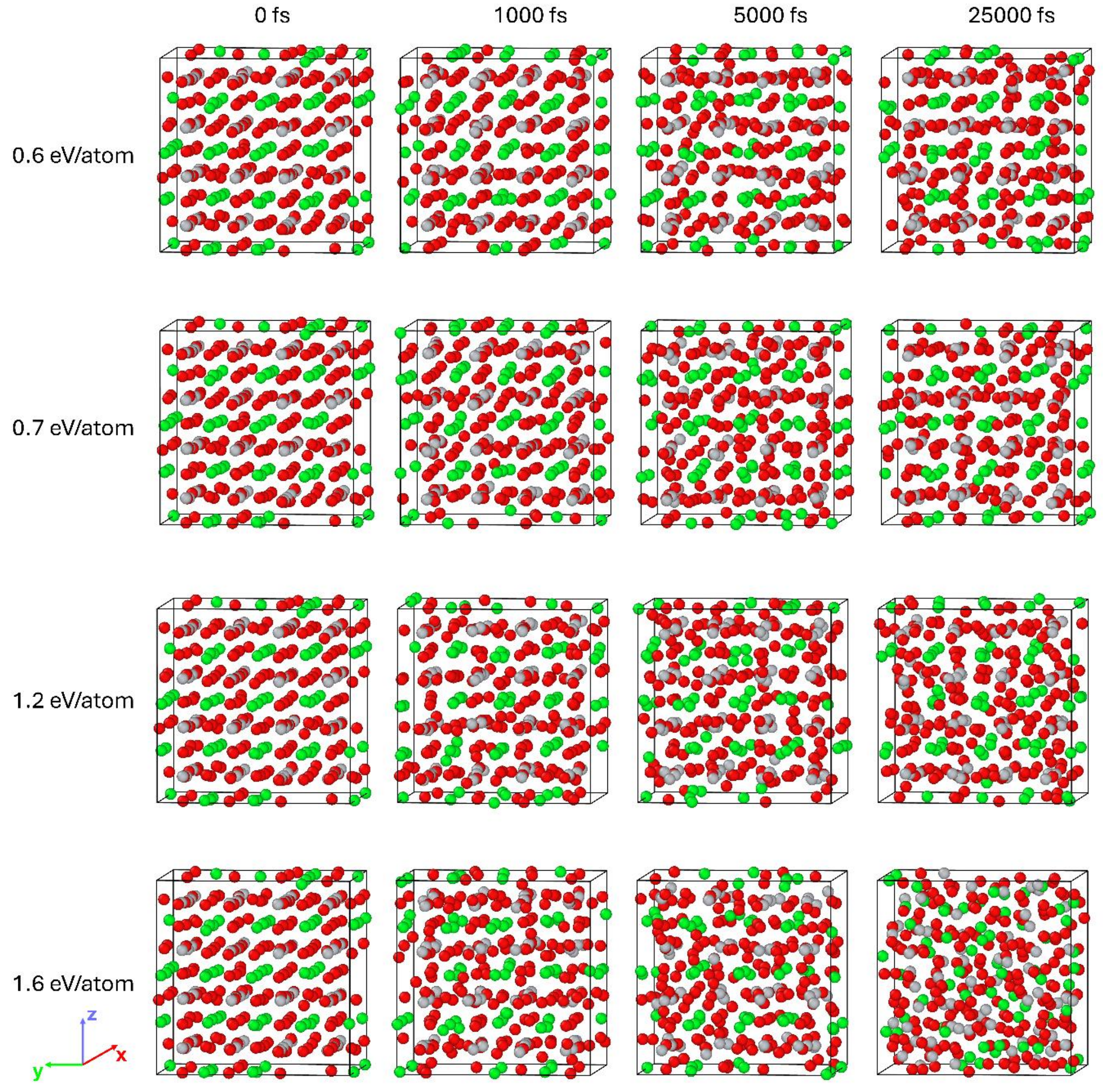

***Figure 16.*** *Atomic snapshots of STO irradiated with a 10-fs Gaussian pulse of 30 eV photon energy at different doses. Red balls are O; green balls are Sr, and grey balls are Ti.*

**Figure 17** shows the orbital population changes alongside the conduction-band electron occupation for all four systems; these data provide a basis for comparison with Hoang et al[95].

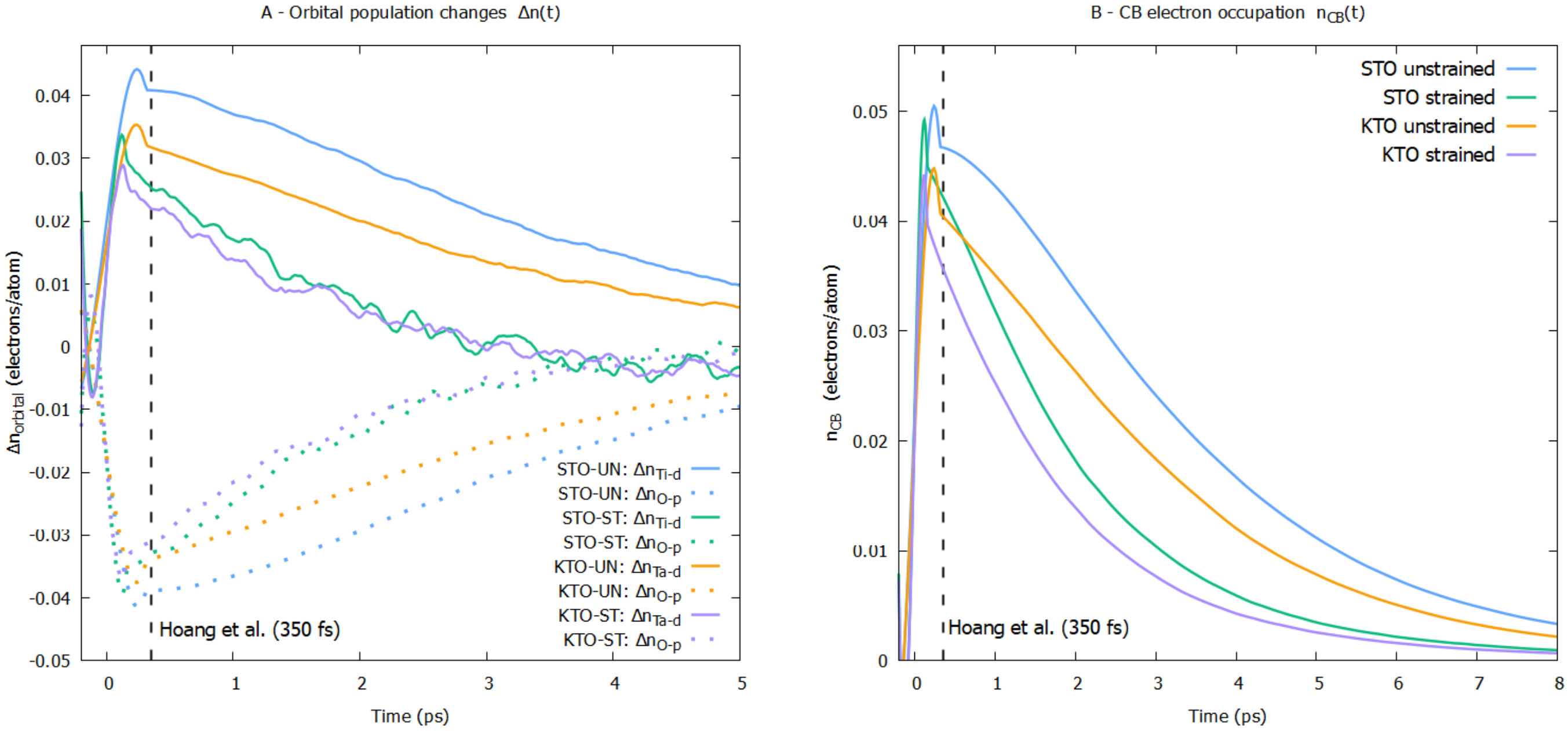

***Figure 17.*** *(Panel A) Orbital population changes* $\Delta n_{B-d}(t)$ *(solid lines) and* $\Delta n_{O-p}(t)$ *(dashed lines), referenced to pre-pulse mean. (Panel B) Conduction-band electron occupation* $n_{CB}(t)$ *for all four systems irradiated with 0.3 eV/atom. In both panels the vertical dashed line at 350 fs marks Hoang et al.*[95] *carrier density peak in* $BaTiO_3$*.*

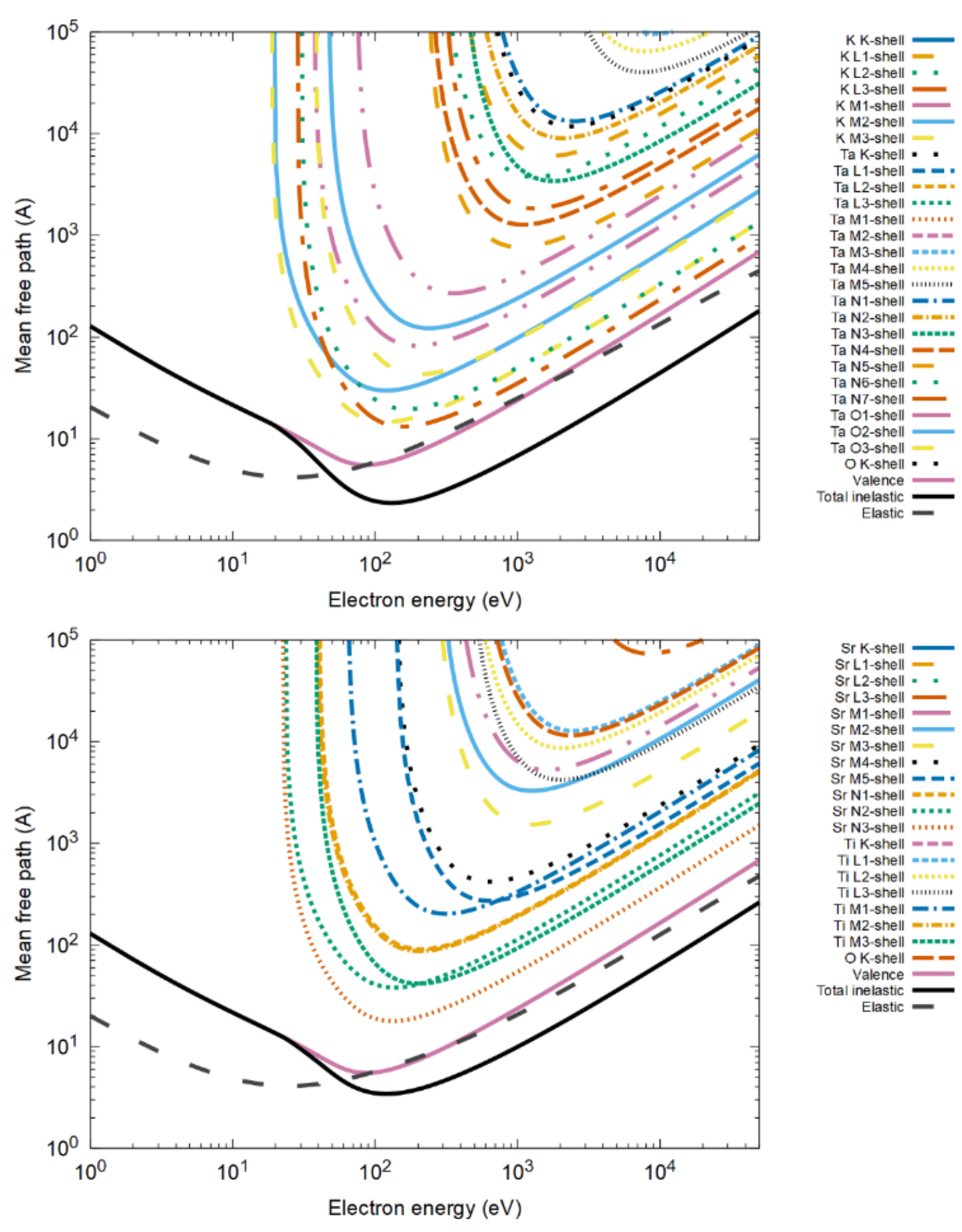

***Figure 18.*** *Element and shell-resolved electron mean free paths in KTO (upper panel) and STO (lower panel).*

**Figure 18** and **Figure 19** document the electron mean free paths and photon attenuation lengths used in XTANT-3 for the calculation of electron kinetics in KTO and STO, respectively.

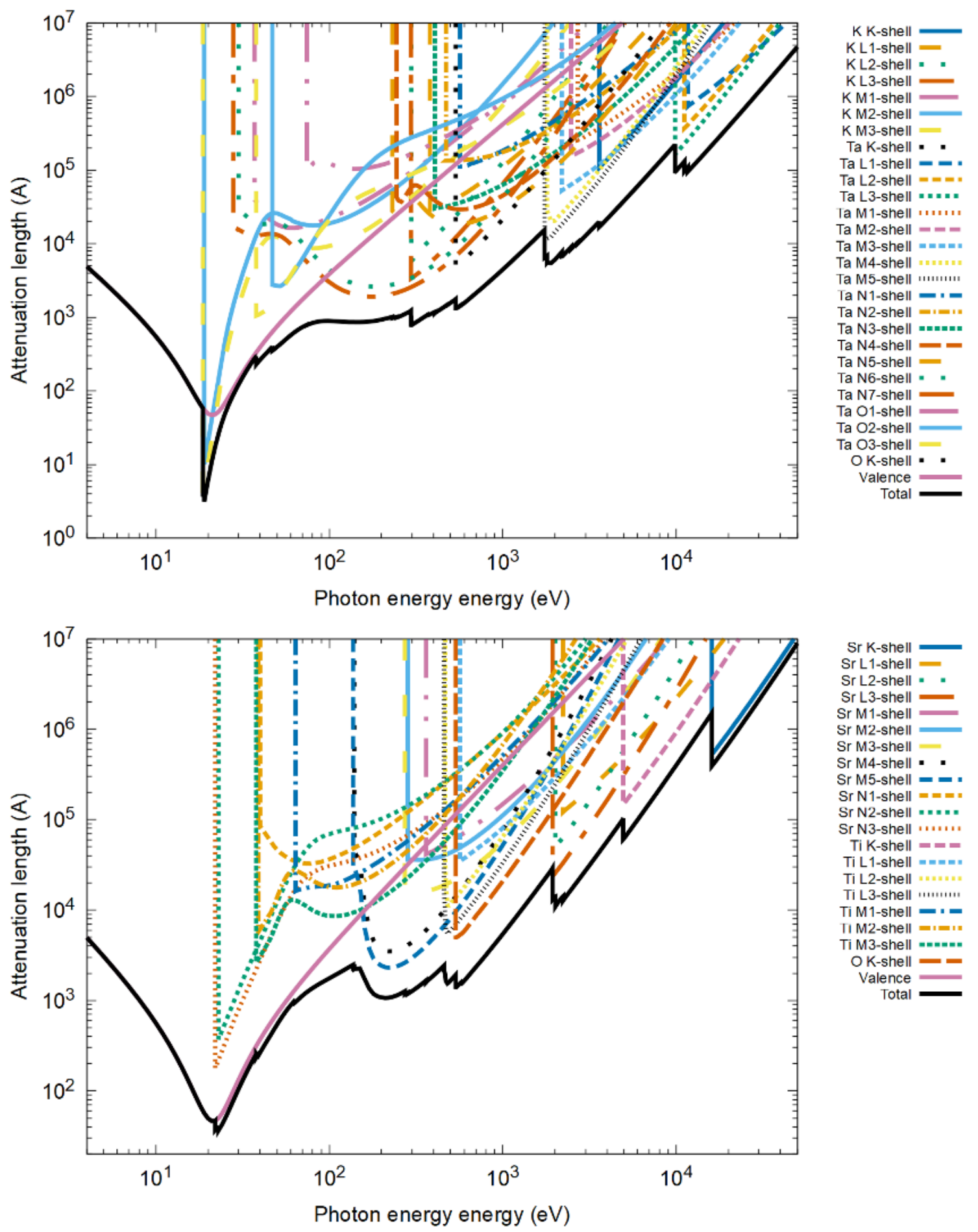

***Figure 19.*** *Element and shell-resolved photon attenuation length in KTO (top) and STO (bottom). Core shell contributions are taken from EPICS2025 while valence contribution is calculated from CDF.*

**Figure 20** and **Figure 21** present powder diffraction patterns, pair correlation functions, and atomic snapshots along crystallographic planes for KTO and STO irradiated at 1.6 eV/atom and cooled with a 10 ps Berendsen thermostat as an approximation to long-term relaxation. The powder diffraction patterns show long-range periodic order in reciprocal space; these may be compared against pump-probe experiments with ultrafast diffraction. The pair correlation function, resolved by atom-pair type, reports the same structural evolution in real space at the level of individual interatomic distances, capturing local coordination changes. The atomic snapshots provide a qualitative view of the same trajectory.

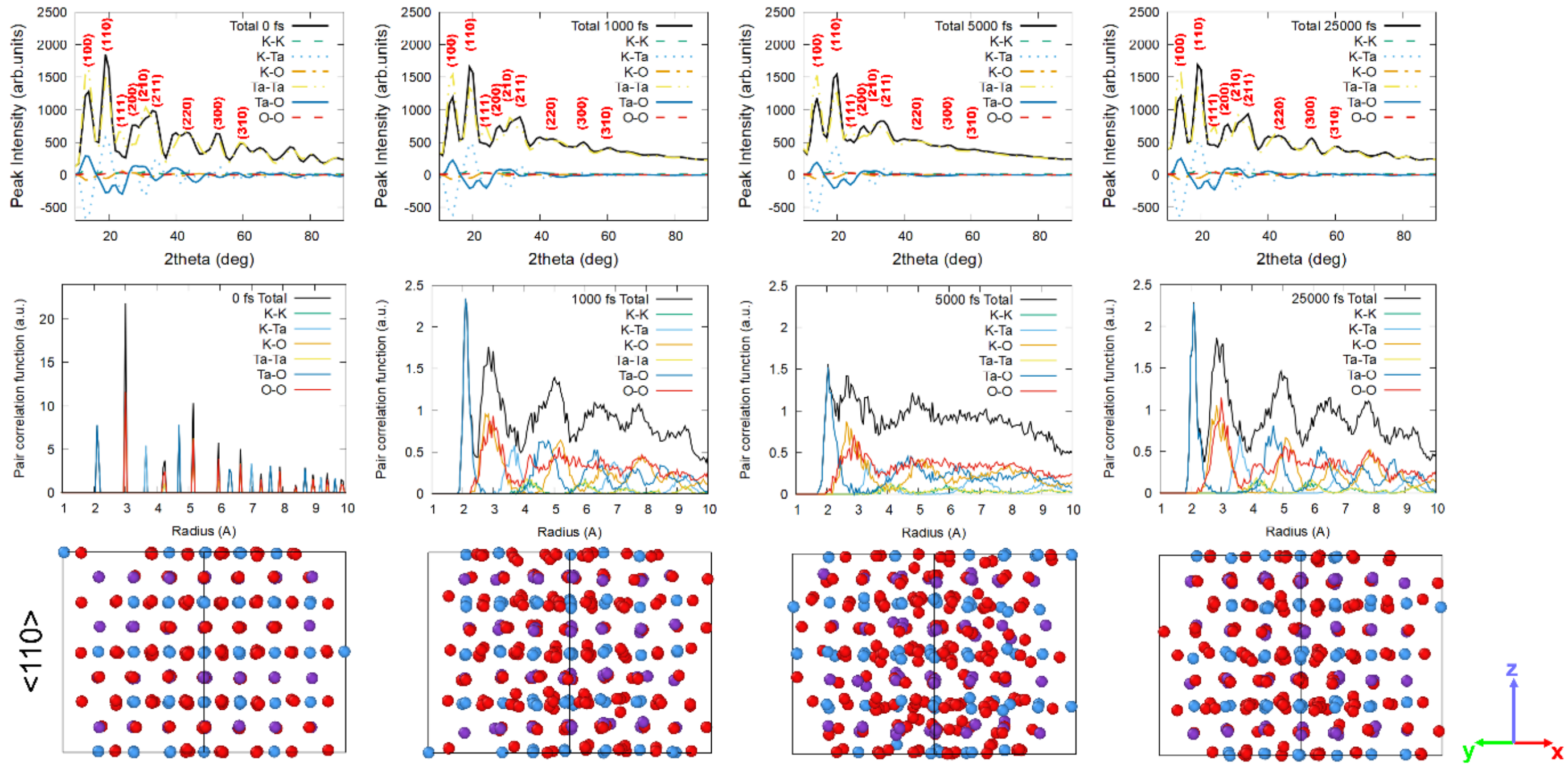

***Figure 20.*** *Powder diffraction patterns (probe wavelength of 1.54 Å, indexes are assigned after validation against experimental reports[99]), pair correlation function and atomic snapshots along the <110> plane in KTO irradiated with a 10-fs Gaussian pulse of 30 eV photon energy at 1.6 eV/atom dose, cooled down with a characteristic time of 10 ps. Red balls are O; violet balls are K; blue balls are Ta.*

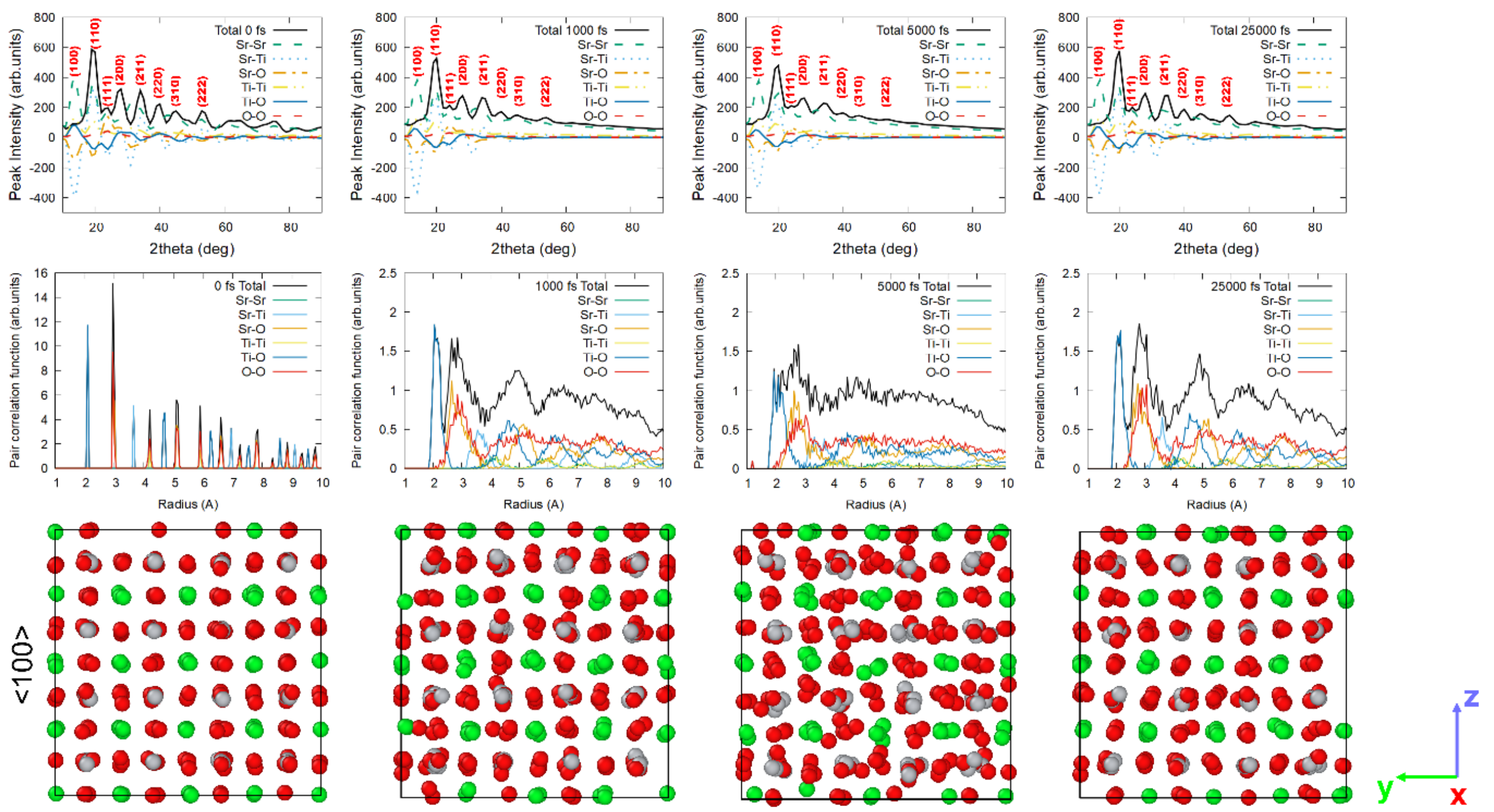

***Figure 21.*** *Powder diffraction patterns (probe wavelength of 1.54 Å, indexes are assigned after validation against experimental reports[100]), pair correlation function and atomic snapshots along the <100> plane in STO irradiated with a 10-fs Gaussian pulse of 30 eV photon energy at 1.6 eV/atom dose, cooled down with a characteristic time of 10 ps. Red balls are O; green balls are Sr, and grey balls are Ti.*